\documentclass[aps,floatfix,prd,showpacs,twocolumn,nofootinbib]{revtex4}
\usepackage{graphicx}
\usepackage{dcolumn}
\usepackage{bm}

\begin{document}

\def\be{\begin{equation}}
\def\ee{\end{equation}}

\title{Modelling DNA Response to THz Radiation}
\author{Eric S. Swanson\footnote{{\tt swansone@pitt.edu}}}
\affiliation{
Department of Physics and Astronomy, 
University of Pittsburgh, 
Pittsburgh, PA 15260, 
USA.}

\date{\today}
\begin{abstract}
Collective response of DNA to THz electric fields is studied in a simple pair bond model.
We confirm, with some caveats, a previous observation of destabilising DNA breather modes and explore the parameter-dependence of these modes. It is shown that breather modes are eliminated under reasonable physical conditions and that thermal effects are significant.
\end{abstract}

\maketitle

\section{Introduction}

There is longstanding speculation that nonionising radiation can damage biological function at the cellular level\cite{early}. More specifically, it has also been suggested that nonionising radiation of varying frequency causes cancer\cite{cancer}. Since nonionising radiation cannot directly disrupt DNA structure, such genotoxic effects must derive from resonance phenomena driven by external electromagnetic radiation\footnote{It is worth remembering that biological electric noise generates internal fields with strengths up to 0.1 V/m\cite{adair}.}

We shall shortly see that the natural frequency of oscillation of DNA base pair separations is approximately 1 THz, thus THz radiation is of special interest. Furthermore, it is very likely that this is the unique frequency range of relevance to bio-resonance effects in DNA.
Interest in this issue has recently been heightened due to the deployment of full body scanners in airports that employ millimeter wave (typically 30-300 GHz) technology. 

Motivated by these observations, Alexandrov {\it et al.} have examined the effects of coupling an electric field driving force to a model of dsDNA bond dynamics\cite{Alex1}. The resulting model of damped, driven, coupled nonlinear oscillators can naturally exhibit exotic collective behaviour (for similar earlier conclusions see Ref. \cite{old-dna}). For example, the familiar period-doubling approach to chaotic dynamics is present. Of more immediate interest is the discovery of a nonlinear discrete breather mode that arises in response to a specific perturbation of the system. This mode stores energy for very long times and can lead to unbinding effects in dsDNA, with obvious implications for the genotoxicity of THz radiation.

While the results of Alexandrov {\it et al.} are compelling, it is unclear if the model is sufficiently robust to permit application to physically realisable situations (such as body scanners).
In particular, this paper critically examines the choice of parameter values, investigates the effect of including thermal fluctuations, and examines the stability of breather modes in a variety of scenarios. It will be shown that parameter variation can eliminate breather modes entirely, or make them unrealistically strong, that thermal noise completely dominates the external influences of the system, and that it is extremely unlikely that dsDNA denaturing can be induced by experimentally accessible THz radiation.

\section{Model Definition and the AGBUR Breather}

The model of Alexandrov {\it et al.} is based on a model of dsDNA pairing dynamics due to Dauxois {\it et al.} (the PBD model)\cite{dpb}. The PBD model employs a Morse potential to model hydrogen bonding between base pairs (and other effects) and an inter-pair stacking potential. Since the nucleotide bonding interactions are much weaker than those of the phosphate-sugar backbone, the degrees of freedom associated with the backbone are neglected. The model also ignores degrees of freedom associated with the helicoidal structure of dsDNA. The resulting model is described by

\be
m_i \ddot y_i = - U_i'(y_i) - W'(y_{i+1},y_i) - W'(y_i,y_{i-1}).
\label{dpb-model}
\ee
where $y_i/\sqrt{2}$ is the deviation from equilibrium distance of the $i$'th base pair. The Morse potential is given by

\be
U_i(y) = D_i [\exp(-a_i y)-1]^2
\ee
and the stacking potential between consecutive base pairs is modelled as\footnote{There is an obvious error in the definition of $W$ in Ref. \cite{Alex1}. This has been repeated in Ref. \cite{Alex2}.}
\be
W(y_i,y_{i-1}) = \frac{1}{2} k (y_i-y_{i-1})^2\,\left(1 + \rho \exp(-\beta(y_i+y_{i+1})\right).
\ee
In general the parameters can depend on the linked base pairs and hence can be labelled $k_{i,i-1}$, etc.
Properties of this model, including the melting transition, were studied in Refs. \cite{dpb,dpb2}.

Alexandrov {\it et al.} chose to supplement the PBD model with periodic driving and frictional terms to model the interactions of dsDNA with an electric field. They state that the interactions of the base pairs with an external electric field are difficult to model and therefore they assume a simple harmonic driving force. The additional terms are then

\be
-m_i \gamma \dot y_i + A \cos \Omega t.
\label{drive-term}
\ee

Evidently the drag term is  not required to model the interactions with an electric field, however such a term is required to produce collective nonlinear phenomena. Finally, Alexandrov {\it et al.} assumed a homogeneous poly(A) DNA molecule with 64 base pairs.
Parameters employed were $m=300$ amu (this was not specified in Ref. \cite{Alex1} -- I assume the value given in Ref. \cite{dpb}), $D= 0.05$ eV, $a = 4.2$ 1/\AA, $k = 0.025$ eV/\AA$^2$, $\beta = 0.35$ 1/\AA, and $\rho = 2.0$. A relaxation time typical to water of $\gamma = 1.0$/ps was used.

As expected, this system displays complicated nonlinear dynamics. Of particular interest is  a breather mode found by Alexandrov {\it et al.} (which we call the AGBUR breather) under a perturbation specified by

\be
\delta y_i(t) = \delta(t-t_0) \, 0.42 \cos[\frac{\pi}{4}(i-i_0)]\, \theta(-4\leq i \leq 4) ({\rm \AA})
\label{fluc}
\ee
at frequency $\Omega = 2.0$ THz and with a drive force of $A = 144$ pN. The breather was localised to be approximately four base pairs wide and had a maximum amplitude of approximately 0.3 \AA.  The authors note that fluctuations like that of Eq. \ref{fluc} can occur thermally and hence transcription and genotoxic effects can be expected.

\subsection{Breather Characteristics}

In preparation for a detailed examination of these claims, we first seek to reproduce the AGBUR breather. Solutions were obtained via a microcanonical molecular dynamics simulation employing the coupled Runge-Kutta (RK4) algorithm. This proved extremely accurate (with relative deviations in total energy of order $5\cdot 10^{-6}$ over 10 ps) and fast. The Verlet method was also implemented, yielding results in agreement with RK4, although less accurate. We follow Ref. \cite{Alex1} and employ 64 base pairs with periodic boundary conditions.

  A breather mode was found at $\Omega = 1.0$ THz, somewhat smaller than the 2.0 THz employed in Ref. \cite{Alex1}. Although it was similar in shape, the maximum amplitude of this breather was found to be about 4 \AA. Note that a compression of $0.42$ \AA, such as generated by the perturbation of Eq. \ref{fluc},  raises the energy of a single bond pair by approximately 1.2 eV, which represents an enormous insertion of energy. In fact the bond length can then be expected to recoil to very large distances, with damping supplied by the stacking potential. Thus one anticipates that large amplitude breathers, such as found here, are to be expected.

\begin{figure}[ht]
\includegraphics[width=9cm,angle=0]{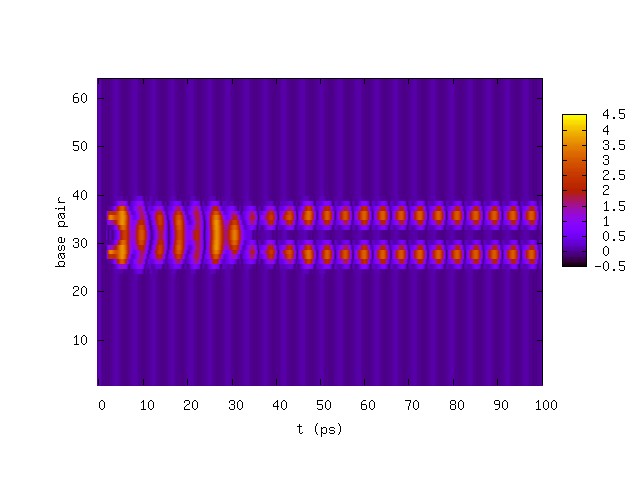}
\caption{(Colour online) A breather mode at $\Omega =1.5$ THz. The compressive perturbation was applied at 2.0 ps. The colour scale represents $y$ (\AA).}
\label{1-fig}
\end{figure}

Another breather with a double-lobe structure was found at a frequency of $\Omega= 1.5$ THz.
This novel mode is shown in Fig. \ref{1-fig}.


One expects that the particular form of the perturbation is not important for the formation of breathers. This has been confirmed by using a compression of the form

\be
\delta y_i(t) = \delta(t-t_0) \, Y_p\, \theta(1 \leq i \leq n).
\label{fluc2}
\ee

\noindent
For $Y_p = -0.42$ \AA, it was found that all perturbations with $n > 4$ generated breathers (with the curious exception of $n=8$). The double-lobe breather was also obtained at $\Omega = 1.5$ THz with this perturbation. At $n=5$ (and $A=144$ pN, $\Omega=1$ THz) one requires a compression of greater than 0.3 \AA\  to achieve a breather mode.

The parameters of Alexandrov {\it et al.} are not the same as those of Dauxois {\it et al.}; in particular the value of $\rho$ was changed from 0.5 to 2.0\cite{Campa}. A run with $n=5$, $Y_p = -0.42$ \AA, and the PBD parameters\footnote{$m = 300$ amu, $D= 0.04$ eV, $a = 4.4.5$ 1/\AA, $k=0.02$ eV/\AA, $\beta=0.35$ 1/\AA, and $\rho = 0.5$.}  reveals that the breather spreads with time, until the entire DNA molecule melts after approximately 140 ps (the same happens with the perturbation of Eq. \ref{fluc}). This is our first indication that breather dynamics are subtle and that model results can depend crucially on parameters.

\begin{figure}[ht]
\includegraphics[width=9cm,angle=0]{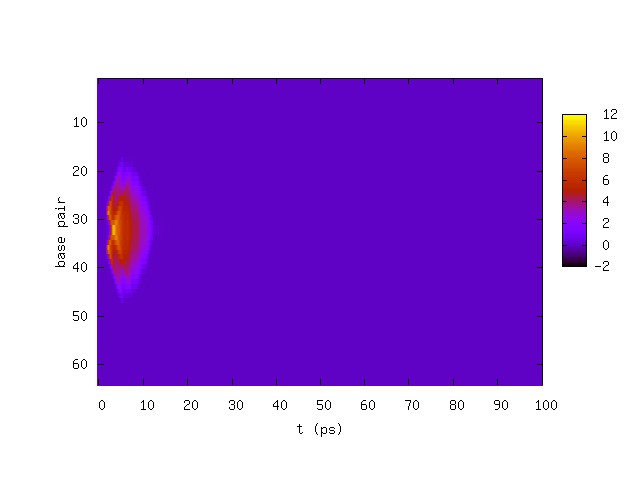}
\caption{(Colour online) System response to a compressive perturbation at 2.0 ps with Barbi parameters. $\Omega = 3.1$ THz. The colour scale represents $y$ (\AA).}
\label{2-fig}
\end{figure}

It should be noted that the parameters of Refs. \cite{Alex1, dpb} are not universally employed. For example, Barbi {\it et al.} have developed a similar model that couples base pair bond extension to helical twist\cite{barbi}. They take $D = 0.15$ eV, $a= 6.3$ 1/\AA, $\beta = 0.5$ 1/\AA, and $k\rho = 0.65$ eV/\AA$^2$. We implement this by assuming $\rho =2.0$ and setting $k = 0.325$ eV/\AA$^2$. Notice that these parameters lead to a considerably stiffer collection of nonlinear oscillators. Indeed, running with the previous drive parameters ($A$, $\Omega$, $Y_p$, $n$) reveals that the breather damps out 
within tens of ps. This remains true for all drive parameters that were tested.  The results make it clear that the relatively large stacking interaction disperses the putative breather (see Fig. \ref{2-fig}).

Finally, the effects of allowing two base pair types are examined. We model this by setting $D= 0.05$ eV for AT pairs and $D=0.075$ eV for GC pairs\cite{ks}; all other parameters are left at their AGBUR values. We find that alternating base pairs or a random configuration of base pairs destabilises the putative breather after approximately 20 ps, again illustrating the fragility the breather mode with respect to parameter variation.

\section{Thermal Effects}

To this point, the genesis of breather modes has relied on the imposition of a perturbative shock (Eqs. \ref{fluc}, \ref{fluc2}) that insert substantial energy into the system. How reasonable are these shocks? Presumably they must be generated by noise within the cell nucleus. This can be due to a variety of biological processes such as cell membrane activity or by simple thermal fluctuations. 

Here we focus on thermal noise and ask the question: how likely is it to perturb a system by $\delta y_i$? Restricting attention to a single base pair, a compression of 0.4 \AA\  corresponds to an insertion of $\Delta U = 1.2$ eV of potential energy to the system. The probability of such a fluctuation is 

\be
p(\Delta U) = \int_{\Delta U}^\infty \rho_{MB}(E) dE \sim \frac{2}{\sqrt{\pi}} \sqrt{x} {\rm e}^{-x}
\ee
where $\rho_{MB}$ is the Maxwell-Boltzmann energy distribution and $x = \Delta U/(k_B T)$.
 The asymptotic form of the error function has been used to obtain this result.

A 0.42 \AA\  compression at room temperature yields $x=46.8$ and a probability $p \approx 10^{-20}$. We now assume that $n$ pair bonds must be compressed, $10^8$ base pairs per dsDNA, $10^2$ dsDNA per cell, $10^{11}$ skin cells\footnote{THz radiation is heavily attenuated and only penetrates 1-2 mm into the body\cite{legal}.}, and a solute collision rate of 1.0/ps, to obtain the estimate

\be
{\cal P} = 10^{-20(n-2)}
\ee
where ${\cal P}$ is the probability of obtaining one breather fluctuation in $n$ base pairs per person per year. We previously established that $n>4$ and hence conclude that such an occurrence is essentially impossible.

Motivated by this result and the observation that the fluctuation-dissipation theorem links friction with thermal noise, we have explored the properties of the model with the addition of Langevin thermal forcing. Thus the term

\be
\eta_i(t)
\ee
has been added to the right hand side of Eq. \ref{dpb-model}. We assume memory-free noise.  The fluctuation-dissipation theorem then implies

\be
\langle \eta_i(t)\eta_j(t')\rangle = 2 m_i \delta_{ij} \gamma k_B T \delta(t-t').
\ee

The molecular dynamics algorithm steps in temporal units $\Delta t$ and therefore we employ the
average noise over a time interval $(t_n,t_n+\Delta t)$:

\be
\bar \eta_i = \frac{1}{\Delta t} \int _{t_n}^{t_n+\Delta t} \eta_i(t) dt.
\ee
Hence
\be
\langle \bar \eta_i^2\rangle = \frac{2 m_i \gamma k_B T }{\Delta t}.
\ee
Thus average noise forces are chosen from a Gaussian distribution

\be
\rho_{th}(\bar \eta) = \frac{1}{\sqrt{2\pi\langle \bar \eta^2\rangle}}\cdot {\rm e}^{-\bar \eta^2/(2\langle \bar \eta^2\rangle)}.
\ee

\begin{figure}[ht]
\includegraphics[width=9cm,angle=0]{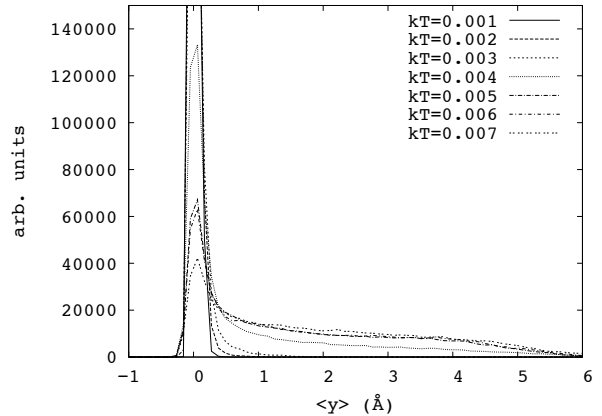}
\caption{Distribution curves for $y$ for various $k_B T$ with AGBUR driving.}
\label{4-fig}
\end{figure}

To test the effect of thermal fluctuations we revert to the homogeneous system with AGBUR parameters and driving forces and eliminate the perturbative shocks of the previous section.  System response was computed at a variety of temperatures.
At low temperature one sees a nonlinear mode with period of approximately 8 ps that is created by the coupled damped  driven oscillators. Increasing temperature leads to rapidly increasing bond length excursions. This can be quantified by plotting the distribution of $y$ versus temperature, as in Fig. \ref{4-fig}. One observes that the distribution is largely invariant for $k_BT \alt 0.003$ eV, $k_B T= 0.004$ is a transition temperature, and temperatures greater than 0.005 eV seem to yield an invariant distribution for $y \agt 1/2$ \AA. Thus it appears the dsDNA with AGBUR driving melts at approximately 0.004 eV (46K).

If one were to take this result seriously it would imply complete chromosomal denaturation in all skin cells in the presence of THz radiation. But the previous section warns of large 
parameter sensitivity in this system and one must not arrive at conclusions too hastily.
In fact using a larger solute relaxation time $\gamma = 2.0$/ps or  increasing the Morse potential strength to $D=0.12$ eV stabilises the system at room temperature.


It should be noted that there is a subtlety concerning the thermal properties of the PBD model. The form of the interaction implies that the equilibrium configuration is two widely separated strands. This is reflected in the partition function, which necessarily diverges. Of course this situation is never realised experimentally because the system is embedded in a complex environment of approximately $\mu$m extent. In our case, the issue does not arise because the molecular dynamics simulation is microcanonical and because it only need be run for hundreds of picoseconds, too short a time scale to probe the asymptotic dynamics of the system. More details concerning the thermal properties of the PBD model can be found in Refs. \cite{Campa,infinity}. I simply add the observation that
Barbi parameters yields tiny fluctuations; in fact the mean bond extension from equilibrium is $\langle y\rangle = 0.03$ \AA\ at 350K, which is inconsistent with the experimentally observed melting transition.

\section{Realistic Driving}

The AGBUR model of driven dsDNA leads to unreasonable results in the presence of thermal fluctuations. We have already seen how parameter variation can alleviate this problem. But another explanation is possible, namely the physical assumptions underlying the model could be inaccurate. We follow this by idea by focussing on the drive term of Eq. \ref{drive-term}. Indeed, this term is of immediate concern, since the authors of Ref. \cite{Alex1} state, ``One complication is that the specific physical nature of the interactions between DNA and the THz electromagnetic field is not known in detail. ... We will here simply augment the PBD [model] to include a drive in the THz frequency range." In fact, I find that the AGBUR breather (at zero temperature, with the AGBUR perturbative shock) requires $A \agt 140$ pN to occur. What is a reasonable physical value for the driving force?

Legal limits on THz radiation power densities range from 5-10 mW/cm$^2$\cite{legal}. The mean magnitude of the Poynting vector $\langle S\rangle = E^2_m/(2 \mu_0 c)$ relates\footnote{The permeabilities of air and water are essentially equal to that of the vacuum.}  this to the maximum electric field strength, $E_m$. Employing the upper limit gives 
\be
E_m \alt 30 {\rm V/m}.
\ee
Assuming that a nucleotide is singly charged then yields a maximum force

\be
A \alt 4 \cdot 10^{-18} {\rm N},
\ee
far smaller than that assumed by AGBUR. In fact the driving force is coupled to the base pair displacement and therefore depends on the nonuniformity of the field over the range $\langle y \rangle$. Assuming an incident plane wave (with wavevector $\vec k$) reduces the strength of the drive force by a factor of $\langle \vec k \cdot \vec y\rangle \sim 10^{-6}$ hence
\be
A \alt 4 \cdot 10^{-24} {\rm N}. 
\ee

But this assumes that the pair bonds all lie in an optimal direction
(the force must be along $\hat y$ but is proportional to $\vec k\cdot \vec y$, thus the pair bond must lie in a plane defined by the wave propagation direction and the direction of electric field oscillation). In reality, DNA is embedded in a heavily hierarchical structure, ranging from the DNA molecule itself to the chromosome. This effectively randomises the pair bond direction with respect to any external field. Thus one should compute $A$ after averaging over bond directions. The result is proportional to contractions of tensors like $\hat E_i \hat k_j \ldots \hat k_\ell$ which is zero to all orders. Thus the driving force relies on remnant order in chromosomal structure and $A$ must be much smaller than $10^{-24}$ N. Finally, approximately one half of the incident radiation is reflected\cite{crc}, cell membranes and cytoplasm are extremely efficient at screening electric fields, even in the THz regime, and the electric charge may be mobile\cite{conduct}. All of these effects reduce the coupling further.
One must conclude that the electric field driving force is many orders of magnitude smaller than that required to generate breather modes.

\section{Discussion and Conclusions}

The PBD model of base pair dynamics is sufficiently rich that interesting collective behaviour can exhibited. Under assumptions concerning drag and drive forcing, breather modes can be generated at certain resonant frequencies. Thus, although this work disagrees on the details, it agrees with the main conclusions of Ref. \cite{Alex1}. The stability of breathers under parameter variation has been addressed here. We have seen that changing $\rho$ from 2.0 to 0.5 or including thermal noise are  sufficient to dissociate dsDNA under AGBUR driving. Alternatively, employing the stiffer Barbi parameters or allowing for a mixture of AT and GC base pairs seems to disallow breather formation.

All of these conclusions are based on drive frequencies near the resonant frequency of the system, a drag term, and a driving term with a magnitude of approximately 100 pN. However, it has been argued that the magnitude of the driving term is much smaller than this. The physical reason is that the source power is rather weak, and DNA is heavily screened from external influences by the cell membrane, the cytoplasm, and the nucleoplasm. The coupling to electric fields is further reduced by the effectively random orientation of a base pair displacement vector. The field strength necessary (estimated generously) to generate 
breather modes is approximately $10^9$ V/m, which is much greater than the dielectric breakdown threshold of air ($\sim 10^6$ V/m).
Thus it appears that the analysis of Refs. \cite{Alex1, Alex2} is not relevant to physically realisable situations.

Although strong THz radiation is artificial, DNA has evolved in a noisy electrical and thermal environment, and it might be expected that the molecule and the processes in which it takes part will be stable 
with respect to external nonionising radiation. Similarly, one would expect that all molecular level biological processes are immune to low intensity nonionising radiation; although, of course, this speculation needs to be confirmed with rigorous experiment.

\acknowledgments
The author is grateful for discussions with D. Boyanovsky, L. Chong, R. Coalson, and D. Jasnow.

\end{document}